\documentstyle[twoside,fleqn,espcrc2,epsfig]{article}
\newcommand{\ec}{\end{center}}

\newcommand{\AmS}{{\protect\the\textfont2
  A\kern-.1667em\lower.5ex\hbox{M}\kern-.125emS}}

\title{New extraction of color-octet NRQCD matrix elements from charmonium 
hadroproduction
\thanks{Work partially
supported by CICYT under contract AEN-96/1718}}
\author{M. A. Sanchis-Lozano$^{a,b}$\thanks{E-mail: mas@evalo1.ific.uv.es}
\vspace{0.4cm}\\
(a) Departamento de F\'{\i}sica Te\'orica \\
\vspace{0.1cm}
(b) Instituto de F\'{\i}sica Corpuscular (IFIC)\\
Centro Mixto Universitat de Val\`encia-CSIC \\
Dr. Moliner 50, E-46100 Burjassot, Valencia (Spain)}

\begin{document}
\begin{abstract}
We re-analyze Tavatron data on charmonium hadroproduction in the
framework of the color-octet model implemented in the event generator
PYTHIA taking into account
initial-state radiation of gluons and Altarelli-Parisi evolution
of final-state gluons fragmenting into $c\overline{c}$ pairs. We 
obtain new values for the color-octet matrix elements relevant
to this production process. We discuss the sensitivity of our results
to the transverse momentum lower cut-off employed in the generation to 
avoid the problematic $p_t{\rightarrow}0$ region, arguing about
the reliability of our previous extraction 
of the NRQCD matrix elements for the $^3S_1^{(8)}$ and
$^1S_0^{(8)}+^3P_J^{(8)}$ contributions.
Finally we extrapolate to LHC energies to get predictions on
the $J/\psi$ direct production rate.
 \end{abstract}
\vspace{0.1in}
\maketitle

\section{INTRODUCTION}
In a series of previous papers \cite{mas0,mas1,mas2,mas3} we analyzed
charmonium hadroproduction in the light of the color-octet model
\cite{braaten}.
We employed a Monte Carlo generator (PYTHIA 5.7) \cite{pythia}
to assess the importance of 
some higher-order QCD effects. As a consequence we concluded that
NRQCD matrix elements \cite{bodwin} determined from Tevatron experimental data
\cite{fermi1}  were considerably overestimated in other analysis
\cite{cho}, since they absorb some
perturbative effects, e.g. initial-state emission of
gluons. The latter gives rise to an effective transverse momentum, 
enhancing the
high-$p_t$ tail of the differential distribution of charmonium.
Once taken into account, the extracted long-distance
parameters have to be lowered significantly. This is especially
apparent for the linear combination of matrix elements (MEs)
 $^1S_0^{(8)}+^3P_0^{(8)}$ 
where, in fact, a large discrepancy w.r.t. HERA results
on $J/\psi$ photoproduction has been pointed out \cite{hera}.\par
However, in our former work we did not consider 
Altarelli-Parisi (AP) evolution of the fragmenting gluon into
charmonium for
the $^3S_1^{(8)}$ contribution at high $p_t$. In this paper we implement
such an effect benefitting from the evolution performed by
PYTHIA itself as will be explained below. We shall show
that although our previous numerical values for the
$^3S_1^{(8)}$ ME were somewhat low, no significant
change occurs for the corresponding $^1S_0^{(8)}+^3P_0^{(8)}$ ME.
We also discuss the sensitivity to the lower cut-off
used to avoid the singular $p_t{\rightarrow}0$ region.
We get started with the latter point in the following section. 

\section{SENSITIVITY TO THE $p_t^{min}$ CUT-OFF}
Production cross sections which are singular at
$p_t=0$ are  automatically regularized in running PYTHIA
by setting a $p_t$ lower cut-off at 1 GeV \cite{pythia}. 
Consistently with our aim of
using by-default options whenever possible in the generation, we
kept this value in all our previous analysis \cite{mas0,mas3}.
\par
In fact, the differential cross section can have a singular
behaviour at low $p_t$ even to next-to-leading order due to 
imperfect cancellation of real and virtual
contributions. However, from a NLO calculation of charmonium 
production by Mangano \cite{mangano}, one may
conclude that our previously used
lower cut-off ($p_t^{min}=1$ GeV) is consistent
with the threshold obtained by Mangano (${\approx}\ 1$ GeV for
the $^1S_0^{(8)}$ channel), such that the averaged cross section is zero 
below this value.

Nevertheless, in order to assess the significance
of the cut-off on the values of the long-distance
parameters we performed new fits to
Tevatron data by changing $p_t^{min}$. We found that, by
varying $p_t^{min}$ in the $[1,2]$ GeV range, the $^3S_1^{(8)}$
ME showed almost no appreciable change, whereas
the $^1S_0^{(8)}+^3P_0^{(8)}$ ME changed by a 
factor two at most \cite{seminar}.\par
Therefore, we keep in our subsequent analysis
the value  $p_t^{min}=1$ GeV as a self-consistent lower
cut-off.

\section{ALTARELLI-PARISI EVOLUTION}
At high transverse momentum, gluon fragmentation via the color
octet mechanism becomes the dominant source of charmonium
production. On the other hand, AP evolution of the 
splitting gluon produces a depletion of its energy which has to be taken
into account. If not so, the long-distance parameter
for the  $^3S_1^{(8)}$ channel would be underestimated.
\par 
In this work the AP evolution of the
fragmenting gluon was carried out from the evolution of
the gluonic partner of the $J/\psi$ in the final state
of the production channel 
\begin{equation}
g\ g\ {\rightarrow}\ g^{\ast}({\rightarrow}J/\psi)\ g
\end{equation}
the technical reason being that the splitting gluon $g^{\ast}$
actually is not generated in our code \cite{mas2}. (Notice
however that the MEs used in the generation include the 
long-distance evolution of the ($c\overline{c}$) bound state 
\cite{mas2}.)

Indeed, on the average the fragmenting gluon should evolve in the 
transverse plane similarly to the other final-state gluon in (1), 
once corrected its virtuality  to become of the order of the 
charmonium mass.

Thereby we get a correcting factor to be applied event by event
to the transverse momentum of the generated
$J/\psi$ (for the $^3S_1^{(8)}$ channel only):
\begin{equation}
x_p\ =\ \frac{\sqrt{P_t^{{\ast}2}+m_{J/\psi}^2}}
{\sqrt{P_t^{2}+m_{J/\psi}^2}} 
\end{equation}
where $P_t$ ($P_t^{\ast}$) is the transverse momentum of 
the final-state gluon without (with) AP evolution.

At high $p_t$,
\begin{equation}
 p_t^{AP}\ =\ x_p\ {\times}\ p_t
\end{equation}
where $p_t$ is the transverse momentum of the $J/\psi$
as generated by PYTHIA (i.e. without AP evolution).\par
Although the above way to implement AP evolution may be
somewhat indirect, it remains in the spirit of our whole
analysis, i.e. using PYTHIA algorithms whenever possible. In fact it
provides an energy depletion of the fragmenting gluon
in accordance with Cho and Leibovich's work \cite{cho}.

\begin{figure}[htb]
\begin{center}
\mbox{\epsfig{file=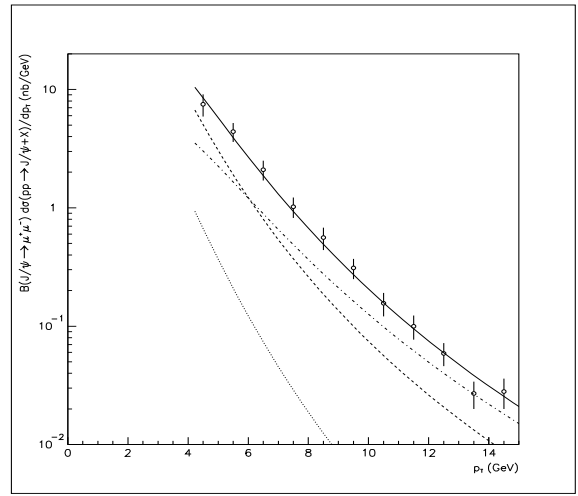,height=6.5cm,width=7.5cm}}\end{center}
\caption{Curves obtained from PYTHIA for $J/\psi$ production
at the Tevatron, including AP evolution and
initial-state radiation on for the CTEQ 2L parton distribution function.
The charm mass was set equal to $1.48$ GeV.
(i) dotted line: Color-singlet model, (ii) dashed line: 
$^1S_0^{(8)}+^3P_0^{(8)}$, (iii) dot-dashed line: $^3S_1^{(8)}$, 
(iv) solid line: all contributions.}
\end{figure}

\begin{figure}[htb]
\begin{center}
\mbox{\epsfig{file=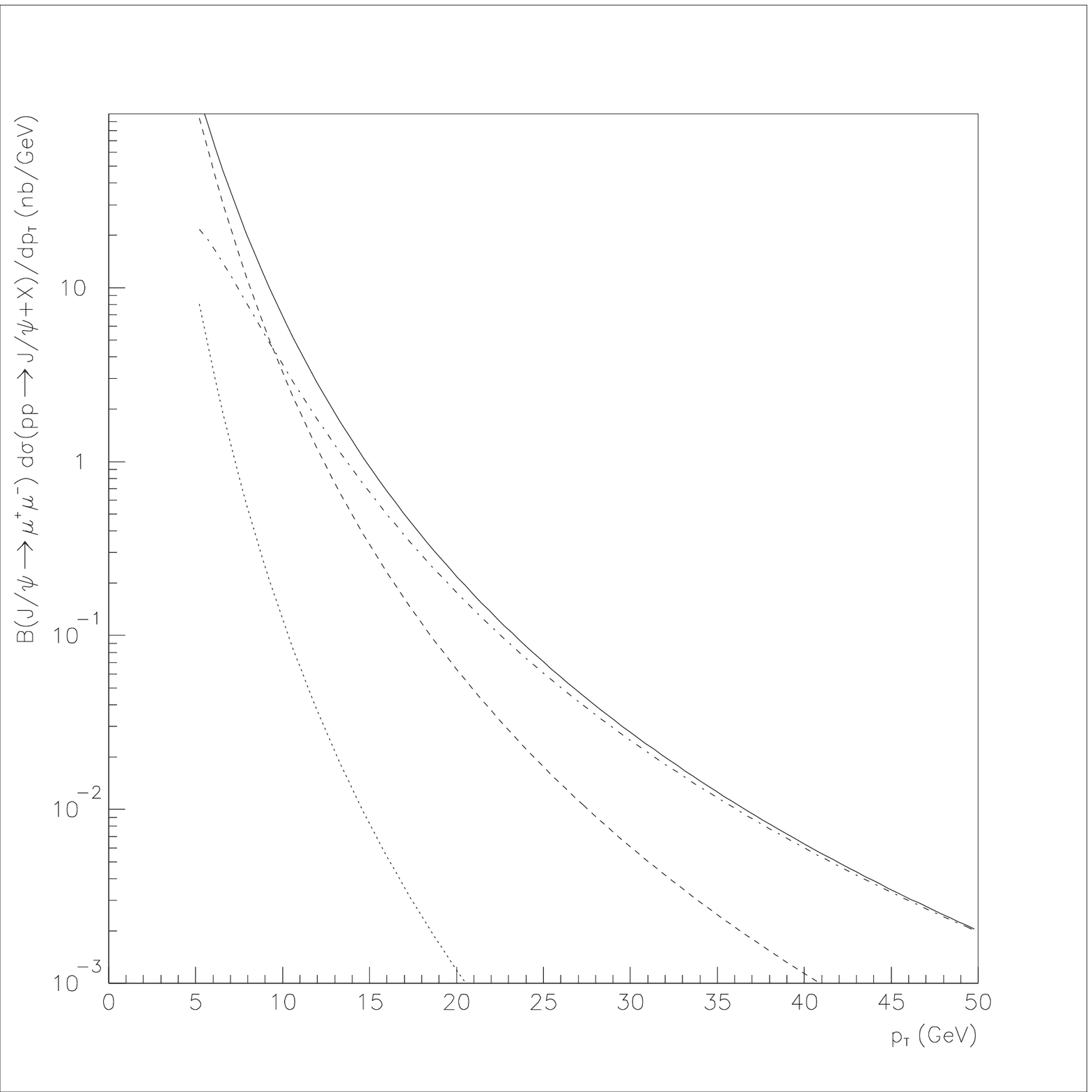,height=6.5cm,width=7.5cm}}
\end{center}
\caption{Curves obtained from PYTHIA for $J/\psi$ production at
the LHC, without AP evolution and
initial-state radiation on for the CTEQ 2L parton distribution function.
(i) dotted line: Color-singlet model, (ii) dashed line: 
$^1S_0^{(8)}+^3P_0^{(8)}$, (iii) dot-dashed line: $^3S_1^{(8)}$, 
(iv) solid line: all contributions.}
\end{figure}

\begin{figure}[htb]
\begin{center}
\mbox{\epsfig{file=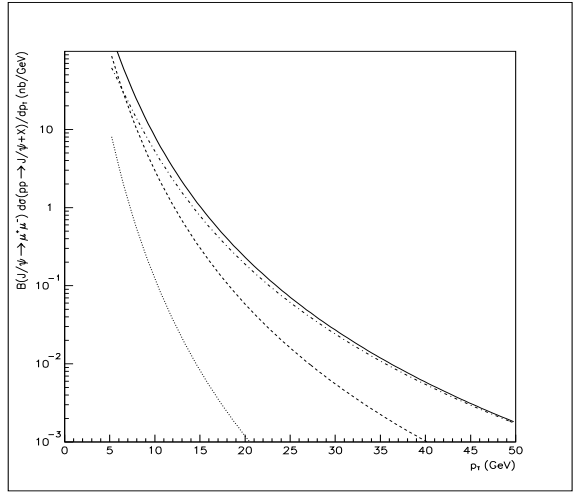,height=6.5cm,width=7.5cm}}
\end{center}
\caption{Curves obtained from PYTHIA for $J/\psi$ production
at the LHC, including AP evolution and
initial-state radiation on for the CTEQ 2L parton distribution function.
(i) dotted line: Color-singlet model, (ii) dashed line: 
$^1S_0^{(8)}+^3P_0^{(8)}$, (iii) dot-dashed line: $^3S_1^{(8)}$, 
(iv) solid line: all contributions.}
\end{figure}

\section{NEW EXTRACTION OF COLOR-OCTET MEs}
We have performed new fits to the Tevatron data \cite{fermi1}
using three PDFs (as in \cite{mas2}) but now 
with AP evolution implemented in the generation as described in the
preceding section. In table 1 we show the numerical values for
 $<O_8(^3S_1)>$ and the linear combination 
$M_3=3{\times}(\frac{<O_8(^3P_0)>}{m_c^2}+
\frac{<O_8(^1S_0)>}{3})$ obtained in our analysis.
The new values for the  $<O_8(^3S_1)>$
matrix elements have been increased by a factor about 3 w.r.t.
AP off \cite{mas2}. Conversely
the $M_3$ value is slightly decreased, moreover remaining quite
smaller than in other analysis \cite{cho}. Therefore we conclude that
the $M_3$ value obtained in our work is reliable and in a
better agreement with HERA results on $J/\psi$ photoproduction \cite{kniehl}.

\begin{table*}[hbt]
\setlength{\tabcolsep}{1.5pc}
\caption{Color-octet matrix elements (in units of $10^{-3}$ GeV$^3$) from 
the best fit to Tevatron data on direct $J/\psi$ production
for different PDFs. Initial-state radiation of gluons and AP evolution as
explained in the text were
swichted on. For 
comparison we quote the values given by Cho and Leibovich: 
$(6.6{\pm}2.1)$ and $(66{\pm}15)$  
respectively.}
\label{FACTORES}

\begin{center}
\begin{tabular}{lcc}    \hline
ME:  & $<O_8(^3S_1)>$ & 
$M_3=3{\times}\biggl(\frac{<O_8(^3P_0)>}{m_c^2}+
\frac{<O_8(^1S_0)>}{3}\biggr)$ \\
\hline

CTEQ2L & $9.6{\pm}1.5$ & $13.2{\pm}2.1$ \\
MRSD0 & $6.8{\pm}1.6$ & $13.2{\pm}2.1$ \\
GRVHO& $9.2{\pm}1.1$ & $4.5{\pm}0.9$ \\
\hline
\end{tabular}
\end{center}
\end{table*}

\section{$J/\psi$ DIRECT PRODUCTION RATE AT THE LHC}
In order to get an estimate of the expected charmonium
production rate at the LHC, we ran PYTHIA with the color-octet
model implemented in. In figure 3 we show the predicted
differential cross section times the muonic branching fraction
of $J/\psi$ with AP evolution included. In figure 2 we show
the same curve with AP off. By comparison, one may conclude that, 
once adjusted
the long-distance parameters to fit Tevatron data, there is no
significant difference between both generations.\par
Let us stress that because of theoretical uncertainties 
associated to different choices for the charm mass, energy scales,
parton distribution function ...,  
the theoretical curves of figures 2 and 3
 have to be considered as order-of-magnitude
predictions. Nevertheless the relatively high production rate
at high-$p_t$ (of the order of the pb at $p_t=50$ GeV) makes especially
interesting the analysis of charmonia production at the LHC. A
preliminary study on $\Upsilon(1S)$ prompt production 
leads to a similar conclusion for the bottomonia family.

\section{CONCLUSIONS}
After checking our analysis by varying the lower $p_t$ cut-off
set in the Monte Carlo generation to avoid the
problematic $p_t{\rightarrow}0$ region, we conclude that our
former choice $p_t^{min}=1$ GeV was justified and our
study of charmonium hadroproduction trustworthy. 
On the other hand, once implemented Altarelli-Parisi
evolution in our framework, the  new extracted values of
the color-octet ME  $<O_8(^3S_1)>$ increase (table 1), recovering
a similar result as in previous extractions \cite{cho}. On the
other hand, the numerical values obtained for 
$M_3$ even decrease slightly, reinforcing the conclusions reported in our
former work \cite{mas0,mas1,mas2,mas3}. 

With respect to the extrapolation of the
$J/\psi$ production rate up to LHC energy, the theoretical
prediction incorporating AP evolution (figure 3) does not differ 
significantly from the
curve obtained without AP evolution (figure 2). This means that, from 
a practical point of view, there is no need to modify the code
for a fast generation as shown in \cite{mas2} if the corresponding
color-octet matrix elements are duly employed. 
Finally we conclude that the analysis of charmonia (and bottomonia) 
prompt production at the LHC deserves special attention in its own right.

\subsubsection*{Acknowledgments}
I thank the LHC Workshop b-production subgroup for interesting
discussions and suggestions.

\end{document}